# Direct Measurement of Periodic Electric Forces in Liquids

B. J. Rodriguez,[1,2] S. Jesse,[1] K. Seal,[2] A. P. Baddorf,[2] and S. V. Kalinin[1,2,*]

[1]Materials Science and Technology Division and [2]The Center for Nanophase Materials Sciences, Oak Ridge National Laboratory, Oak Ridge, TN 37831

The electric forces acting on an atomic force microscope tip in solution have been measured using a microelectrochemical cell formed by two periodically biased electrodes. The forces were measured as a function of lift height and bias amplitude and frequency, providing insight into electrostatic interactions in liquids. Real-space mapping of the vertical and lateral components of electrostatic forces acting on the tip from the deflection and torsion of the cantilever is demonstrated. This method enables direct probing of electrostatic and convective forces involved in electrophoretic and dielectroforetic self-assembly and electrical tweezer operation in liquid environments.

---

[*] Corresponding author, sergei2@ornl.gov

Electric forces in solution play an important role in a wide range of systems including polyelectrolytes, electrical double layers, charged lipid membranes, and biomolecular systems.[1,2] Furthermore, a number of methods in biology and biotechnology, including electrophoresis and dielectrophoresis, directly utilize electrically controlled molecular and particle motion in solutions. In the last decade, applications of microelectrodes to trap and manipulate, e.g., cells and viruses at the microscale,[3,4] and routes for micro- and nanofabrication through electrophoretic[5] and electrostatic[6] self-assembly have been demonstrated. Finally, a number of microfluidic and scanning probe microscopy-based techniques are being developed to allow electrical control of single cells and molecules in solution using electric forces (electric tweezers). These applications require a quantitative understanding of electrical interactions in liquids on the nanometer scale. In ambient and ultra high vacuum environment, electrostatic interactions are dominated by parabolic (in bias) non-dissipative capacitive forces.[7] In liquid, additional effects due to the presence of double layers, electrochemical reactions, mobile ions, ionic currents, and convective motion, etc. of the liquid must be taken into account.[8,9] This complexity of interactions necessitates experimental methods to probe electrically-driven force interactions in liquids.

Previously, the effect of electrostatic forces in liquids was extensively studied using measurements based on direct static force detection and force-distance spectroscopy.[10,11,12] In these, the electric charge density is controlled through the partial dissociation of surface chemical groups (i.e., is controlled by the pH of the solution). Recently, we demonstrated that an ac signal can be applied to an atomic force microscope (AFM) tip in solution to measure the local electromechanical response of a ferroelectric sample.[13] Furthermore, the application of a dc bias in certain solvents could be used to induce ferroelectric switching, providing information on dc electric field localization in the solution.[14] Here, we demonstrate an approach similar to scanning impedance microscopy (SIM)[15,16] to measure the electric force between two periodically biased electrodes in solution.

Measurements are performed using an Asylum Research MFP-3D AFM with an additional function generator and lock-in amplifier (DS 345 and SRS 844, Stanford Research Systems). The scanner head was modified to allow direct access to the tip deflection and torsion signals after the first-stage amplifier, bypassing the microscope electronics. The system was equipped with an external LABVIEW/MATLAB data acquisition system for frequency spectroscopy. Measurements were performed using a standard tip-holder with Pt coated tips (Electri-Lever, $l \approx 240$ μm, resonant frequency ~ 70 kHz, $k$ ~ 2 N/m) in the dual-pass mode. In the dual pass mode, the cantilever is driven mechanically at the first resonance during the main line. During the second pass, the tip retraces the topography of the surface at a specified lift height and the torsion or deflection of the tip at the frequency of electrical modulation are recorded to yield electric force signals. Similar results were obtained in the single pass mode, where the tip was mechanically driven at the first resonance and electrically driven at the second. To obtain quantitative values for the electric force, the cantilever spring constants and static deflection sensitivities were calibrated as described elsewhere.[17,18,19]

The interdigitated metal electrode sample was fabricated using standard photolithographic techniques. The sample was mounted on an ORCA sample holder with conducting clips on the contact pads. An ac-bias was applied between the electrodes, and 1kΩ current limiting resistors were connected in series, as described elsewhere.[15] During the measurements, the sample was kept floating and the tip was biased or kept at ground. For liquid imaging, a drop of deionized (DI) water was placed on the sample. Additional water was added as needed to compensate for evaporation.

The topography and electric force amplitude and phase maps acquired in liquid at a 50 nm lift height are shown in Fig. 1. Tip deflection and torsion maps are acquired sequentially between the same electrode pair. The amplitude map for the tip torsion (Fig. 1(b)) reveals that the lateral forces are strongest in the gap between the biased electrodes. At the same time, the flexural force is maximized over the biased electrode (Fig. 1(e)). This behavior is anticipated from simple electrostatic considerations.

In Fig. 2, the voltage dependence of the vertical electric force is shown in air (Figs. 2(a) and 2(b)) and in liquid (Fig. 2(c)). In Fig. 2(a), the tip is biased with 3 V, while the tip is ground in Figs. 2(b) and 2(c). In air, with a biased tip, the signal on the biased electrode is linear in voltage, as expected. Without tip-bias, the signal for both biased and ground electrodes scale similarly with a linear dependence above a certain threshold. In DI water, the signal from the ground electrode is null, while the signal from the biased electrode scales with ac voltage similarly to the trend in air with a ground tip. Note that operation in DI water requires the tip potential to be small to avoid electrochemical reactions. Also in Fig. 2, the lift height dependence of the vertical electric force is shown in air (Figs. 2(d) and 2(e)) and in liquid (Fig. 2(f)). In all cases, the signal maxima are close to the surface.

In Fig. 3, the torsional electric force on the cantilever as a function of voltage and lift height are shown in air with a biased tip and in liquid with a ground tip. The ac bias dependence of the amplitude of the tip torsion is shown in Figs. 3(b) and 3(c) for air and DI water, respectively. In air there is a linear dependence, while in liquid the dependence is more complex and follows the same trend as the tip deflection. The trends are also similar for the torsional signal as a function of lift height.

Finally, the frequency dependence of the electric force is shown in Fig. 4. Here, tip calibration has been used to determine the force quantitatively (below the first resonance). The slow-axis scan was disabled allowing the frequency of the bias to the electrodes to be changed at the end of each scan line. Maps of the amplitude of the deflection signal as a function of position and frequency in air and DI water are shown in Fig. 4 (a,d), respectively. The averaged deflection amplitude signal as a function of frequency is shown in Fig 4 (c). Maps of the torsional amplitude signal as a function of position and frequency in air and DI water are shown in Fig. 4 (b,e), respectively. The averaged torsional amplitude signal as a function of frequency is shown in Fig 4 (f). In air, the effective electrical force deflecting the cantilever is on the order of $F_{eff}$ = 225 nN near the first resonance. The absolute force is $F = F_{eff}/Q$, where Q is the *Q*-factor, and is 1.8 nN. For the second resonance, this force is 0.1 nN. In liquid, this force is 0.8 nN. This suggests the macroscopic convective motion of the liquid induced by the periodic electrode bias dominates the probe dynamics.

To summarize, we have demonstrated direct measurement of electrostatic forces in liquid using biased electrode array. In DI water, the electrostatic signal is detectable on the length scale of ~100 nanometers, comparable to estimated Debye length. The dc tip control is limited in aqueous environment due to the onset of electrochemical reactions. The non-linear effects and rectification at the electrode-solution interface also affect the bias dependence of the signal, resulting in non-linear behavior. Despite these limitations, this approach provides a direct measure of electrophoretic (first harmonic) and convective forces due to the periodic bias applied to the electrodes. Potentially, dielectrophoretic interactions can be determined from the second harmonic component of forces acting on a tip, Hence, this method opens the pathway for direct probing of electric interactions in liquids. Furthermore, it may allow local bio-electric fields to be measured near cells and proteins. The further development of the technique will include the development of the insulated probes to minimize tip-electrode stray currents, and isolated microelectrochemical cells to minimize exposed electrode area.

# Figure Captions

**Fig. 1.** (a,d) ac-mode height, images of the ac-biased electrodes in DI water, and (b,e) amplitude and (c,f) phase images of tip torsion and deflection signals, respectively.

**Fig. 2.** Amplitude signal of the tip deflection as a function of ac-bias (a) in air with a biased tip, (b) in air with a ground tip, and (c) in DI water with a ground tip. Amplitude signal of the tip deflection as a function of lift height (d) in air with a biased tip, (e) in air with a ground tip, and (f) in DI water with a ground tip.

**Fig. 3.** (a) ac-mode height, amplitude and phase images of the ac-biased electrodes in DI water, and tip torsion amplitude and phase images at 3 different lift heights. Amplitude signal of the tip torsion as a function of ac-bias (b) in air with a biased tip and (c) in DI water with a ground tip. Amplitude signal of the tip torsion as a function of lift height (d) in air with a biased tip and (e) in DI water with a ground tip.

**Fig. 4.** Maps of the amplitude of the (a,d) deflection and (b,e) torsion signals as a function of position and frequency in air and DI water, respectively for a 12 micron slow-axis-disabled scan. (c,f) The averaged amplitude signal as a function of frequency for tip deflection and torsion, respectively. The frequency range in liquid is limited by the onset of electrochemical processes.

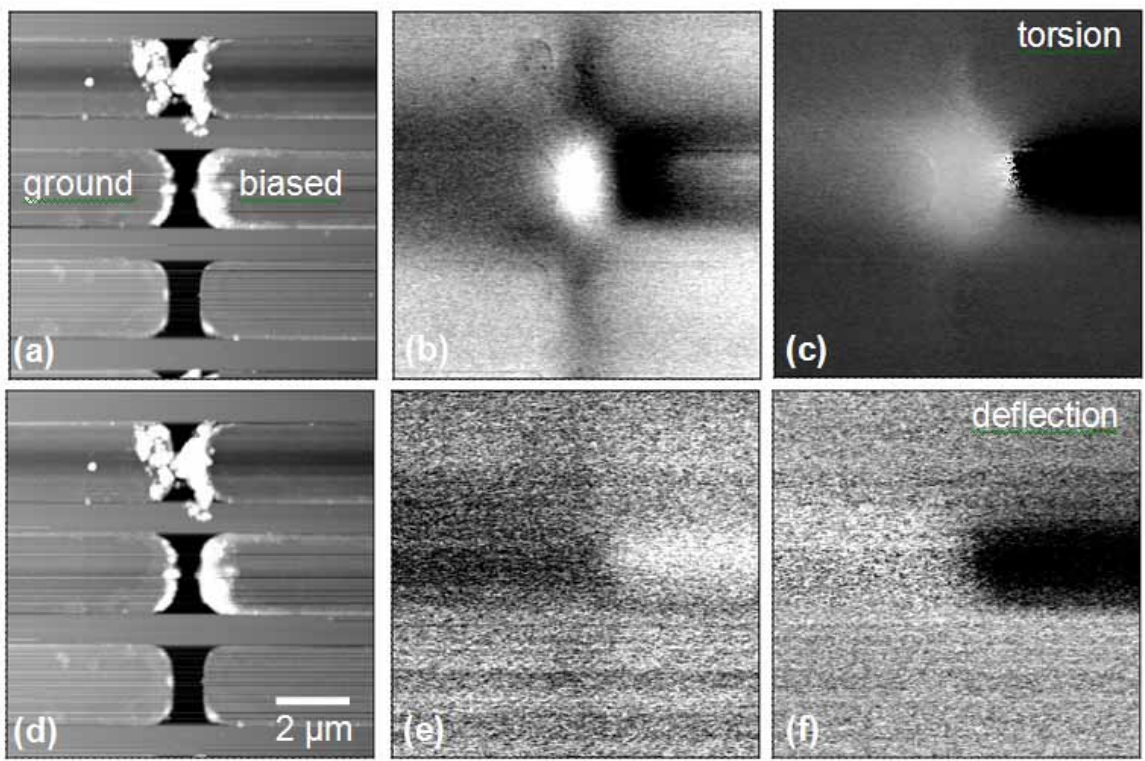

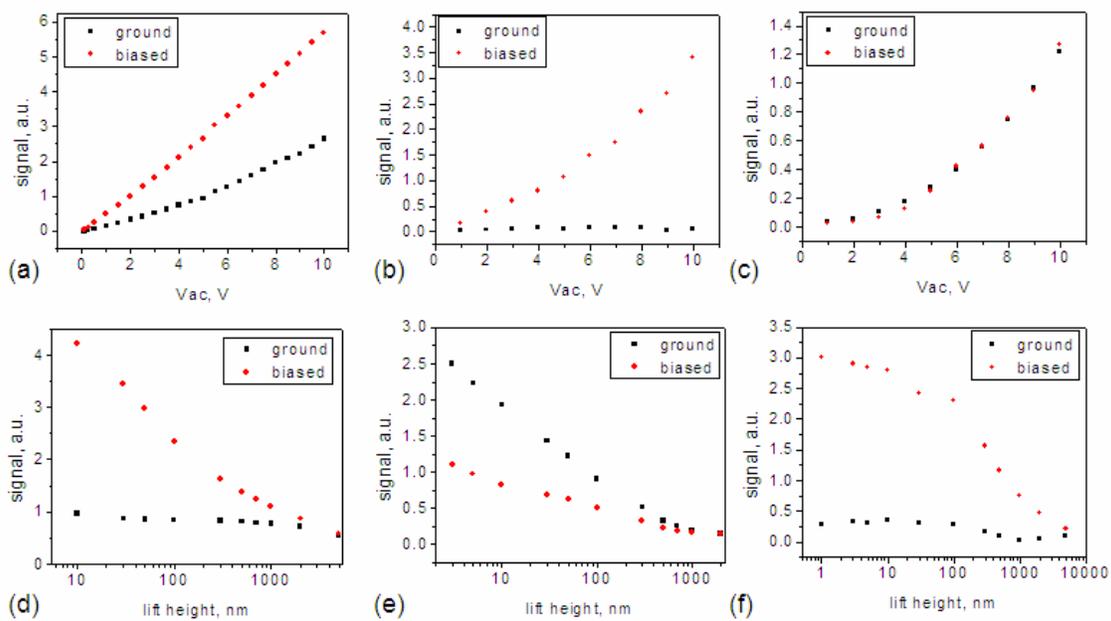

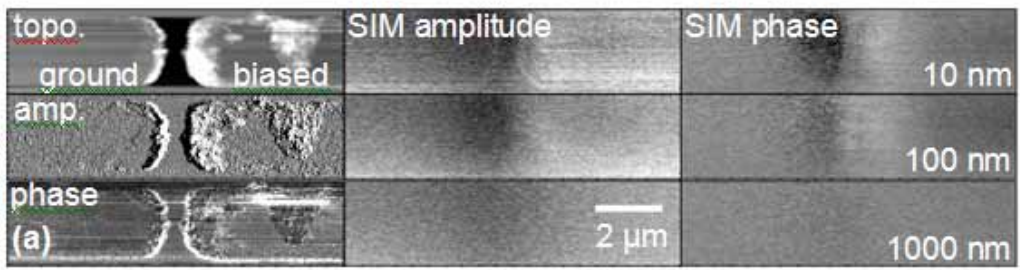
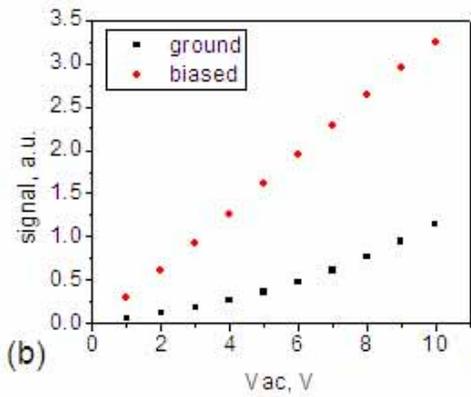
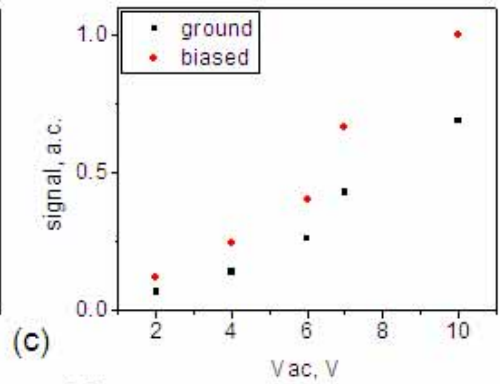
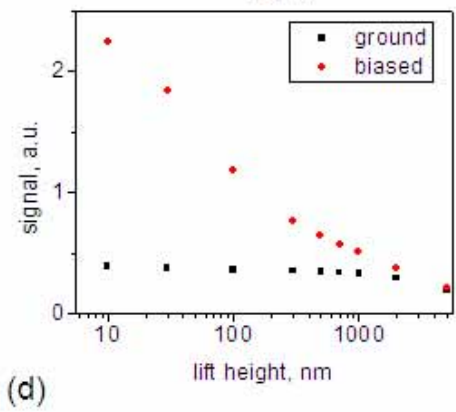
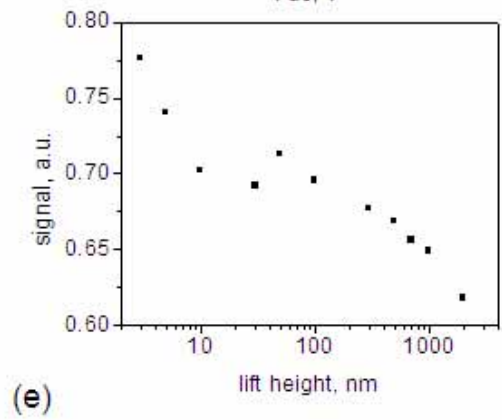

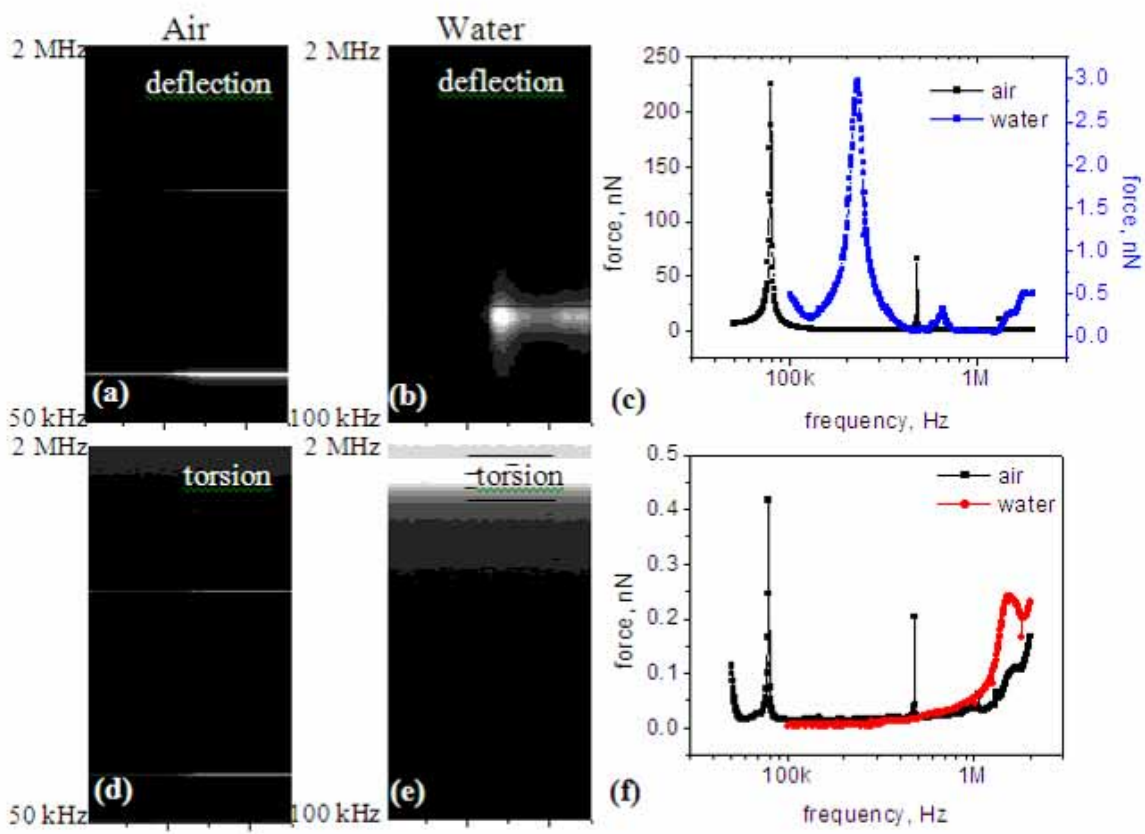